\documentclass[aps,pra,twocolumn,groupedaddress]{revtex4}
\usepackage{graphicx}

\begin{document}


\title{Quasi-2D Confinement of a BEC in a Combined Optical and Magnetic Potential}


\author{N. L. Smith, W. H. Heathcote, G.Hechenblaikner, E. Nugent and C. J. Foot}
\affiliation{Clarendon Laboratory, Department of Physics, University of Oxford,\\
Parks Road, Oxford, OX1 3PU, \\
United Kingdom.}


\date{\today}

\begin{abstract}
We have added an optical potential to a conventional Time-averaged
Orbiting Potential (TOP) trap to create a highly anisotropic
hybrid trap for ultracold atoms. Axial confinement is provided by
the optical potential; the maximum frequency currently obtainable
in this direction is 2.2~kHz for rubidium. The radial confinement
is independently controlled by the magnetic trap and can be a
factor of 700 times smaller than in the axial direction. This
large anisotropy is more than sufficient to confine condensates
with $\sim10^{5}$ atoms in a Quasi-2D (Q2D) regime, and we have
verified this by measuring a change in the free expansion of the
condensate; our results agree with a variational model.
\end{abstract}


\maketitle


\section{Introduction}
%
%
Two-dimensional systems are of great interest in condensed matter
physics in general, and they have some special properties
\cite{Binney},\cite{Cardy}. A two-dimensional Bose gas confined in
a uniform potential does not undergo Bose-Einstein condensation
(BEC); instead there is a Berezinskii-Kosterlitz-Thouless (KT)
transition: a topological phase transition mediated by the
spontaneous formation of vortex pairs which leads to a system that
does not have long-range order but which is nevertheless
superfluid. Many features of the KT transition have been observed
in experiments with films of superfluid He$^4$ on surfaces
\cite{Bishop1978}\cite{Kosterlitz1973}; (see \cite{Luhman} for
some recent results). In experiments with a thin layer of
spin-polarised atomic hydrogen gas on a liquid helium surface
\cite{Safonov1998}, evidence was found for the existence of a
quasi-condensate: a system that only possesses local phase
coherence, in contrast to a pure condensate that has a global
phase.

The recent advent of laser cooling and subsequent
Bose-condensation of the alkali metal atoms \cite{Anderson1995a}
has opened up a new avenue of investigation: while these bosonic
gases are initially cooled to quantum degeneracy as three
dimensional clouds, the addition of novel dipole force potentials
can make these systems two dimensional to varying degrees. A
condensate can be called quasi-two dimensional (Q2D) when the
energy level spacing in one dimension exceeds the interaction
energy between the atoms. In this case the particles obey 2D
statistics but interact in the same way as in a three-dimensional
system.

The first Q2D condensate was made with sodium atoms that were
condensed in a magnetic trap and then loaded into an attractive
optical dipole trap \cite{Gorlitz2001}. The dipole trap was made
by tightly focusing a laser beam along one direction with a
cylindrical lens to achieve trap aspect ratios of 79. In this
experiment the number of atoms, and hence the interaction energy,
was reduced to enter the Q2D regime. A Q2D condensate of two
thousand caesium atoms has also been made in a surface wave trap
with an aspect ratio of 50 \cite{Rychtarik2004}. In our experiment
we use two sheets of blue-detuned light to increase the axial
confinement of our magnetic trap, this approach allows
anisotropies in excess of 700 to be achieved as the radial
confinement is provided independently by the magnetic trap, which
in turn allows more atoms to be loaded into the Q2D regime. The
combination of our magnetic Time-averaged Orbiting Potential (TOP)
trap and dipole potential creates a flexible system with the
ability to rotate the condensate, so that vortex nucleation can
also be studied.

Q2D condensates have many interesting properties that are not
present in 3D: the effect of phase fluctuations has been examined
theoretically in \cite{Petrov2000}; the possibility of vortex
production as a mechanism for the decay of a quadrupole mode is
examined in \cite{Fedichev2003}. Indeed while it has been shown
that in a harmonically trapped non-interacting gas Bose-Einstein
condensation can occur \cite{Bagnato1991}, there is still a
question of when an interacting 2D system supports a KT rather
than a BEC transition.

Our paper is structured as follows: in the second section we
describe the design of our optical potential, in the third section
we discuss its realisation and some initial calibration
measurements, and in the final section we present our measurements
of the condensate in free expansion which confirm it has been
trapped in the Q2D regime.

\section{Dipole Trap Design}

\subsection{Optical Potential}

In order to produce strong axial confinement, we trap the atoms in
the nodal plane at the focus of a Hermite-Gaussian TEM$_{01}$-like
mode. To produce this light field the output from a laser at
wavelength $\lambda=532$~nm (\emph{Coherent Verdi}) is passed
through a phase-plate, which imprints the top half of the beam
with a $\pi$ phase shift relative to the bottom half. The
phase-plate was manufactured in house, and coated in such a way as
to balance the reflective losses from the upper and lower
surfaces. After focusing with a cylindrical lens with focal length
$f=160 ~\text{mm}$ in the $z$ direction, the intensity pattern at
the focus resembles a TEM$_{01}$ mode to a first approximation.

\begin{figure}
\includegraphics[width=\columnwidth]{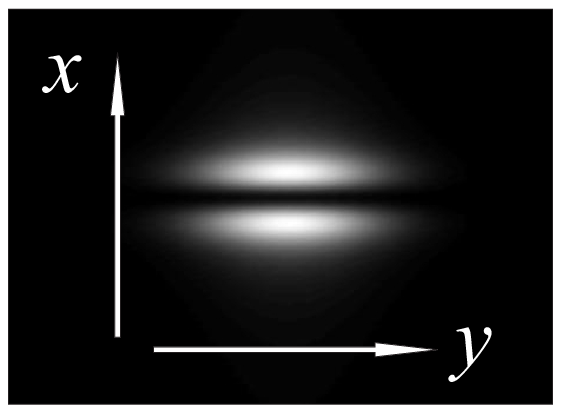}\\
\caption{TEM$_{01}$-like intensity distribution at the focus of
the final cylindrical lens. Axis convention is shown: $z$ is in
the vertical direction, $y$ is horizontal across the beam and $x$
is along the direction of beam propagation, out of the page. The
$y$ and $z$ directions are not to scale. The radial direction is
to be interpreted as in the $xy$ plane.\label{intensityandaxes}}
\end{figure}

\begin{figure}
\includegraphics[width=\columnwidth]{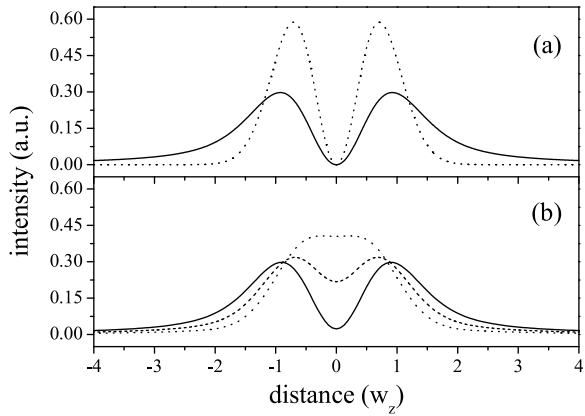}\\
\caption{(a) The intensity distribution in the focal plane for the
actual experimental setup (solid line) and for the ideal
TEM$_{01}$ profile (dotted line) of the same total power. (b) The
actual potential is plotted for various values of the displacement
parameter: $\xi=0.3$ (solid line), $\xi=1.0$ (dashed line),
$\xi=1.5$ (dotted line).\label{plotpotentials}}
\end{figure}

We derive the exact expression for the intensity distribution
below and also discuss any shortcomings together with the effects
of various types of misalignment. The axis convention used in this
paper is explained in Fig.~\ref{intensityandaxes} and the
coordinate axes are denoted by capital letters for the input plane
(at the position of the cylindrical lens) whereas they are denoted
by small letters in the focal plane, where the atoms are trapped.
The intensity distribution of the initial beam is given by
\begin{equation}
I_{\mathrm{init}}(Y,Z)=\frac{2 P}{\sigma_y \sigma_z \pi}
\exp{\left(-\frac{2Y^{2}}{\sigma_{y}^{2}}
\right)}\exp{\left(-\frac{2Z^{2}}{\sigma_{z}^{2}} \right)},
\end{equation}
where $\sigma_y,\sigma_z$ are the beam waists in $Y$ and $Z$
direction, respectively, and $P$ is the power in the laser beam.
The amplitude function $\tilde{\phi}(y,z)$ in the focal plane is
given by
\begin{equation}
\tilde{\phi}(y,z)=\frac{1}{\sqrt{\lambda f}}
\int_{-\infty}^{\infty} \exp\left(\mathrm{i} \tilde{k}Z\right)
\phi(y,Z) dZ,\label{tildephi}
\end{equation}
where $\phi(y,Z)=\phi(Y,Z)=\sqrt{I_{init}(Y,Z)}~\gamma(Z-Z_0)$ is
the amplitude function in the input plane. The function
$\gamma(Z-Z_0)=1$ for $Z<Z_0$ and $\gamma(Z-Z_0)=-1$ for $Z>Z_0$.
It represents the action of the phase-plate which, as we shall
assume in the following, is aligned with the center of the
Gaussian beam for $Z=Z_0$. The cylindrical lens performs a 1-d
Fourier transform and $\tilde{k}=kz/f$, $k=2\pi/\lambda$. After
some calculations we obtain for the intensity in the focal plane
\begin{eqnarray}
&&I_{\mathrm{foc}}(y,z)=\left|\tilde{\phi}(y,z)\right|^2 \label{erfiintensity}\\
&&=\frac{2 P}{\lambda
f}\frac{\sigma_y}{\sigma_z}\text{erfi}\left(\frac{k \sigma_z
z}{2f}\right)^2 \exp\left(-\frac{k^2 \sigma_z^2 z^2}{2 f^2}\right)
\exp\left(-\frac{2y^2}{\sigma_y^2}\right),\nonumber
\end{eqnarray}
where the imaginary error-function $\text{erfi}(z)$ is defined in
terms of the conventional error function $\text{erf}(z)$ evaluated
for a purely imaginary argument:
$\text{erfi}(z)=-\mathrm{i}~\text{erf}(\mathrm{i}z)$. To first
order $\text{erfi}(z)\approx (2/\sqrt{\pi}) z+O(z^3)$. Inserting
this approximation to $\text{erfi}$ into the above equation and
expanding it up to second order in $z$ we obtain
\begin{equation}
I_{\mathrm{foc}}(z)=\frac{8 \pi P\sigma_z^3}{\lambda^3f^3\sigma_y}
z^2 +O(z^4). \label{frequencies}
\end{equation}
Provided that $\Delta \gg \Delta_{FS}$ \cite{Grimm1999a}, where
$\Delta$ is the detuning of the dipole trap from resonance with
the atomic transition and $\Delta_{FS}$ is the fine structure
splitting of the transition, the potential energy of an atom in
the dipole trap is related to the intensity of the light field by
\begin{equation}
U_{\mathrm{dip}}(\textbf{r}) = \frac{3\pi
c^{2}}{2\omega_{0}^{3}}\frac{\Gamma}{\Delta}I(\textbf{r}),\label{potential}
\end{equation}
where $c$ is the speed of light \emph{in vacuo}, $\Gamma$ and
$\omega_{0}$ are the linewidth and frequency of the atomic
transition, and $I(\underline{r})$ is the intensity of the dipole
trapping beam. In our experiment the atomic transition is the
$^{2}S_{1/2}\rightarrow^{2}P_{1/2,3/2}$ in atomic Rb$^{87}$ which
has a wavelength of 780~nm. From
Eqs.(\ref{frequencies},\ref{potential}) we obtain for the axial
trap frequency $\Omega$ for atoms of mass $m$,
\begin{equation}
\Omega^2=\frac{24\pi^2 P \sigma_z^3 c^2
\Gamma}{\lambda^3f^3\sigma_y\omega_0^3 m\Delta}.\label{Omega1}
\end{equation}
Going back to Eq.(\ref{erfiintensity}) we shall define the beam
waists in the focal plane as $w_y=\sigma_y$ and
$w_z=2f/(k\sigma_z)$. The latter relation is exactly that between
the input and the output waists of a Gaussian beam. A Gaussian
beam of any order transforms into different sized Gaussian beam of
the same order through the action of a lens. In our case the phase
jump introduced by the phase-plate causes a zeroth order Gaussian
to transform into something closely resembling a first order
Gaussian beam (TEM$_{01}$) rather than a zeroth order. Using the
expressions for the beam waists in the focal plane the intensity
profile (\ref{erfiintensity}) can be written as
\begin{eqnarray}
&&I_{\mathrm{foc}}(y,z)=\nonumber\\
&&\frac{2 P}{\pi w_yw_z}\text{erfi}\left(\frac{z}{w_z}\right)^2
\exp\left(-\frac{2 z^2}{w_z^2}\right)
\exp\left(-\frac{2y^2}{w_y^2}\right).\label{Ifocus}
\end{eqnarray}
The axial trap frequency $\Omega$ can then be expressed in terms
of these waists as
\begin{equation}
\Omega^{2} = \frac{24c^{2}\Gamma P}{\pi m\omega_{0}^{3}
w_{z}^{3}w_{y}\Delta}. \label{Omega2}
\end{equation}
In order to determine the trap frequency $\Omega$ for our
experimental parameters we can measure the TEM$_{00}$ profile in
the input plane and then use Eq.(\ref{Omega1}) to calculate the
resulting trap frequencies of the optical potential in the focal
plane. Alternatively, we can measure the distance $d_z$ between
the two intensity peaks in the focal plane and use the relation
$d_z=1.8483~w_z$ to determine $\Omega$ from Eq.(\ref{Omega2}). The
beam intensity profile, given by Eq.(\ref{erfiintensity}) differs
somewhat from that of a perfect TEM$_{01}$ which is given by
\begin{equation}
I_{\mathrm{TEM_{01}}}(y,z) = \frac{8 P}{\pi w_z^3w_y}
z^{2}\exp{\left(-\frac{2z^{2}}{w_{z}^{2}}
\right)}\exp{\left(-\frac{2y^{2}}{w_{y}^{2}} \right)}.
\label{tem01}
\end{equation}
Substituting $\text{erfi}(z)\approx 2/(\sqrt{\pi})z$ into
Eq.(\ref{Ifocus}) we obtain Eq.(\ref{tem01}) with the initial
multiplying constant smaller by a factor of $\pi$. This
approximation is only valid for small $z$ but it indicates that in
the central region, where we want to confine the atoms, the
intensity of the focused beam is reduced by a factor of $\pi$ with
respect to that of a pure TEM$_{01}$ beam of the same power $P$.
The reason is that much power is lost in the very broad wings of
the focused beam which are much larger than those of a pure
TEM$_{01}$. This can be seen in Fig.\ref{plotpotentials}a, where
the solid line shows the actual potential given by
Eq.(\ref{Ifocus}) and the dotted line shows the potential of a
TEM$_{01}$ given by Eq.(\ref{tem01}) for comparison.

The initial amplitude distribution has a sudden jump in the middle
where the amplitude goes from negative to positive. The focusing
lens performs a Fourier transform and in order to resolve the
sudden change in amplitude many higher spatial frequencies,
represented by the wings of the intensity profile in the focus,
are needed.

Combining the magnetic with the optical potential we find that the
central position of the combined potential, in the axial
direction, is given by

\begin{equation}
z_{c}(t) = \frac{z_{l}}{1 +
\frac{\omega_{m}^{2}}{\omega_{l}^{2}(t)}} \label{zposeqn},
\end{equation}

where $z_{c}$ is the position of the combined potential relative
to the center of the magnetic trap, $z_{l}$ is the position of the
center of the optical potential, and $\omega_{l}$ and $\omega_{m}$
are the angular frequencies of the optical and magnetic potentials
respectively.

\subsection{Beam Shaping and Misalignment Errors}
\begin{figure}
\includegraphics[width=\columnwidth]{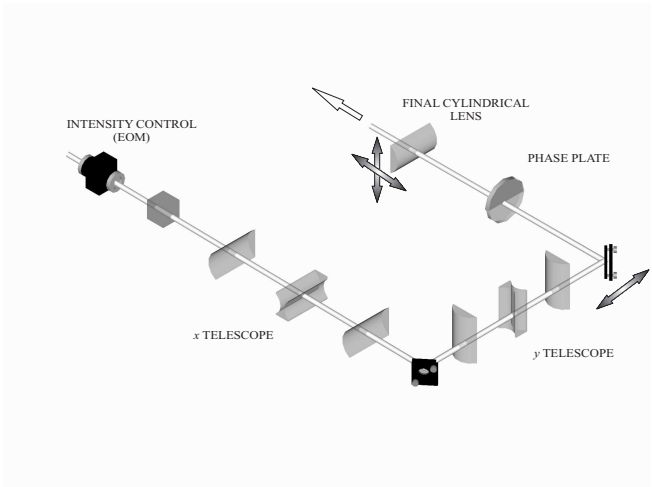}\\
\caption{Schematic drawing of optical layout showing the
cylindrical telescopes, the $\pi$ phase-plate and the final
cylindrical lens. Shaded arrows indicate the location of
sub-micron actuators for positional control over the
focus.\label{schematicopticallayout}}
\end{figure}

A schematic diagram of the experimental setup is shown in
Fig.~\ref{schematicopticallayout}. To achieve a high axial
confinement frequency, a tight focus is required in the $z$
direction, as such the beam is expanded in this direction prior to
focusing. A suitable choice of beam width is also required in the
$y$ direction to determine the area in the $xy$ plane throughout
which Q2D confinement can be realized. The $y$ and $z$ beam widths
are not the same, and so some method of asymmetric beam expansion
is required: for this experiment two orthogonal, cylindrical
telescopes are used. To achieve a beam focus of 7.2~$\mu$m with a
final lens of focal length f=160~mm, the beam waist in the $z$
direction before focusing must be 3.8~mm. The beam waist in the
$y$ direction is chosen to be 410~$\mu$m. From Eq. (\ref{Omega1})
the resulting axial trap frequency is $\approx 2.7 \text{kHz}$.

\begin{figure}
\includegraphics[width=\columnwidth]{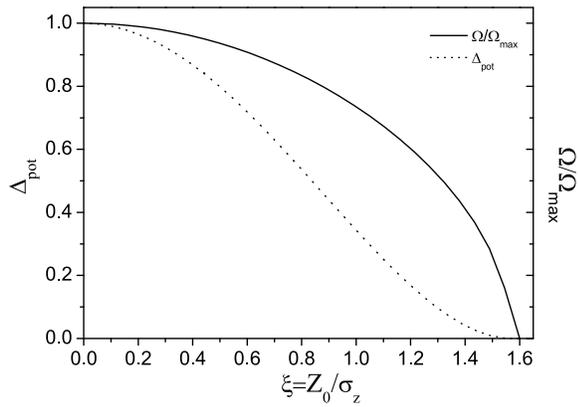}\\
\caption{The oscillation frequency for rubidium atoms in the
potential normalised by its value for $\xi=0$
($\Omega/\Omega_{max}$) is plotted against the phase-plate
displacement parameter $\xi$ (solid line). The potential height
$\Delta_{pot}$ decreases with increasing $\xi$ (dotted
line).\label{freq_change}}
\end{figure}

The potential height and frequencies are decreased by misaligning
the phase-plate with respect to the incoming Gaussian beam. Two
types of misalignment can occur. The first one is displacement of
the phase step on the phase-plate from the center of the Gaussian
beam. In this case $Z_0$, defined below Eq.(\ref{tildephi}), is
not equal to zero. As a result the intensity dip in the central
region is fills up for large values of the displacement parameter
$\xi=Z_0/\sigma_z$. Fig.~\ref{plotpotentials}b shows a series of
potential plots for various values of $\xi$. The central dip has
completely disappeared for $\xi=1.5$. The resulting change in trap
frequencies is shown in Fig.~\ref{freq_change} (solid line) which
plots $\Omega$ normalised by its value when $\xi=0$. The dotted
line plots the potential height normalised by its value for
$\xi=0$:
$\Delta_{pot}=(I_{max}(\xi)-I_{min}(\xi))/(I_{max}(0)-I_{min}(0))$.

\begin{figure}
\includegraphics[width=\columnwidth]{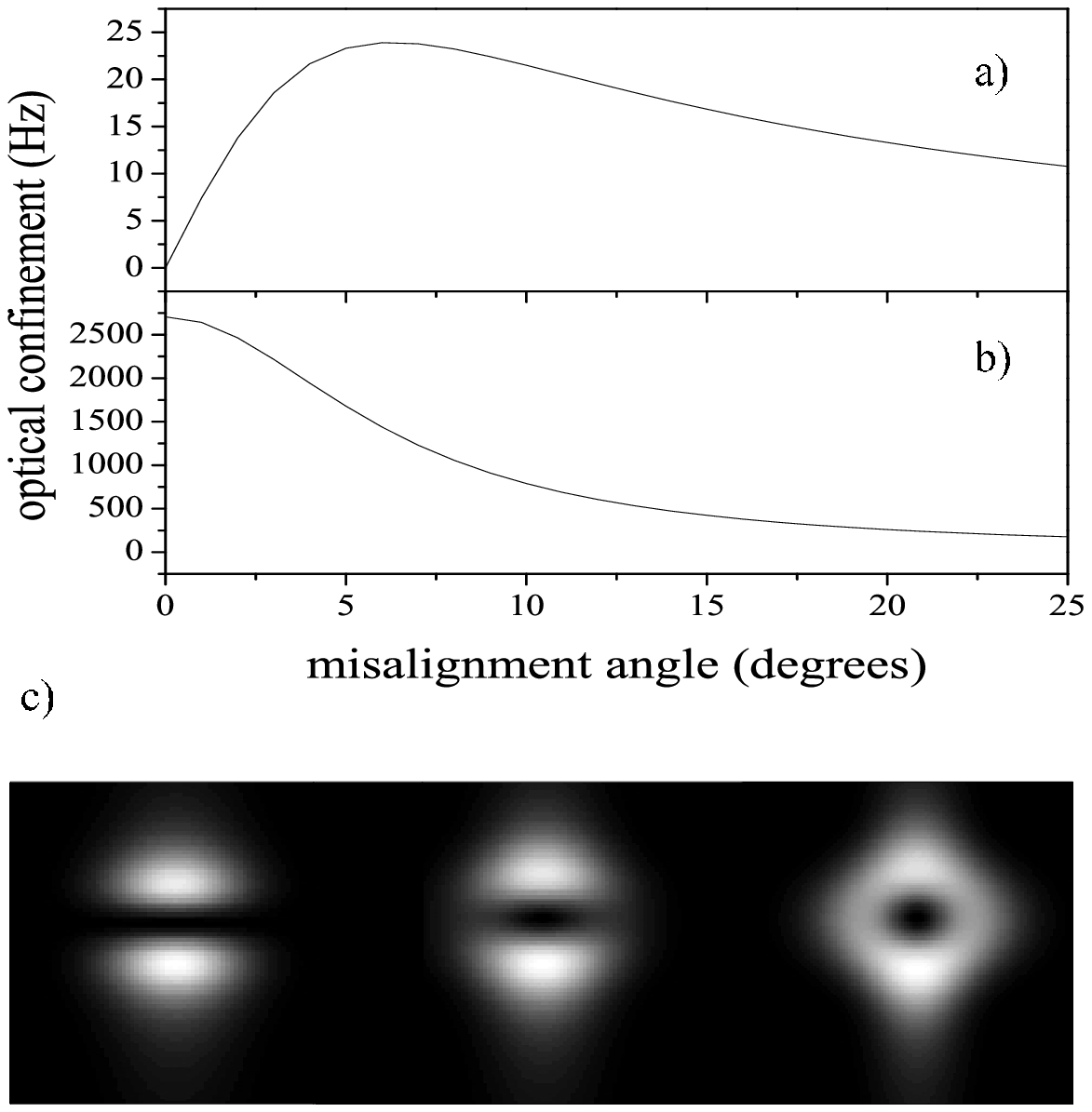}\\
\caption{The effect of angular misalignment between the input beam
and the final lens, calculated for $w_{z}$=7.2~$\mu$m and
$P=4.5$~W. (a) Plot shows the confinement in the $y$ direction as
a function of angular misalignment. (b) Plot shows decrease in
axial trapping frequency as a function of misalignment. (c)
Intensity distributions at the focus of the cylindrical lens for
2, 5 and 10 degrees of input beam misalignment (left to right; in
all the plots the $y$ and $z$ axes are not the same
scale).\label{anglemisaligncombined}}
\end{figure}

The effect of rotational misalignments between the input beam, the
phase-plate and the final cylindrical lens have been calculated
numerically for our experimental parameters (see
Fig.~\ref{anglemisaligncombined}). The results show that the
optical potential at the beam focus is relatively insensitive to
the precise rotational alignment of the phase-plate; however, the
angle that the elongated input beam makes with the cylindrical
lens is much more critical, as misalignment decreases the axial
confinement and introduces some radial confinement. In our
experiment we aim to minimise the radial confinement and maximise
the axial confinement provided by the optical potential.

\section{Experimental Realisation}

\subsection{Alignment}

Sub-micron actuators give independent control over the focus of
the optical potential in three dimensions. The $x$ and $z$
position of the focus can be controlled by moving the final
cylindrical lens, and the $y$ position is changed by moving the
position of a mirror which deflects the beam through 90 degrees
(see Fig.~\ref{schematicopticallayout}). Intensity control is
provided by an Electro-Optic Modulator (EOM).

The size and depth of the optical potential are too small to
provide any noticeable change to the Magneto-Optical Trap (MOT)
used in the first stage of our experiment, and so initial
alignment was carried out by making the beam from a 397 nm
semiconductor diode laser co-linear with the dipole trapping beam.
Photons at this frequency are sufficiently energetic to ionize Rb
atoms and deplete the MOT, providing a clear signature for
alignment. Subsequent coarse alignment was carried out by lining
up the MOT with the entry and exit reflections of the trapping
beam on the glass cell. This allows the beam to be aligned to
within ~200~$\mu$m, in $x$ and $y$, of the magnetic trap center.

For the next stage of alignment the atoms were supported against
gravity on top of the laser beam. The power in the trapping beam
was then reduced, until the atoms were only just supported, and
the beam moved in the $xy$ plane: as the beam focus is moved
closer to the atoms, less power is required to stop the atoms from
falling. This is an iterative process, which is accurate to $\sim
30~\mu m$ in $x$ and $y$.

The most critical direction for alignment in our experiment is the
$z$ direction, which needs to be aligned to within a few microns
to load atoms from the magnetic trapping potential into the dipole
trap. Alignment in this direction exploits a delay that is
required between the switch off of the magnetic and optical
potentials. The dipole trap can be switched off by the EOM in a
few microseconds, however the magnetic trap requires almost
$500~\mu s$ before the current in the quadrupole coils decays.
This can create a problem when looking at the expansion of
condensates that are tightly confined by the optical potential:
simulations show that when suddenly released, the expansion of the
condensate can be slowed or even temporarily reversed by the
residual magnetic fields. The solution is to introduce a delay of
$500~\mu s$ after the magnetic field begins to decay, before
switching off the optical potential. The equilibrium position of
the atoms in the combined potential is not quite the same as the
equilibrium position in the purely optical potential: the result
is that the atoms move in the optical potential during the delay.
The velocity they acquire during this period is sufficient after
15~ms of free expansion to allow the start position to be
calculated, and is accurate to $\sim 1~\mu m$.

\begin{figure}
\includegraphics[width=\textwidth]{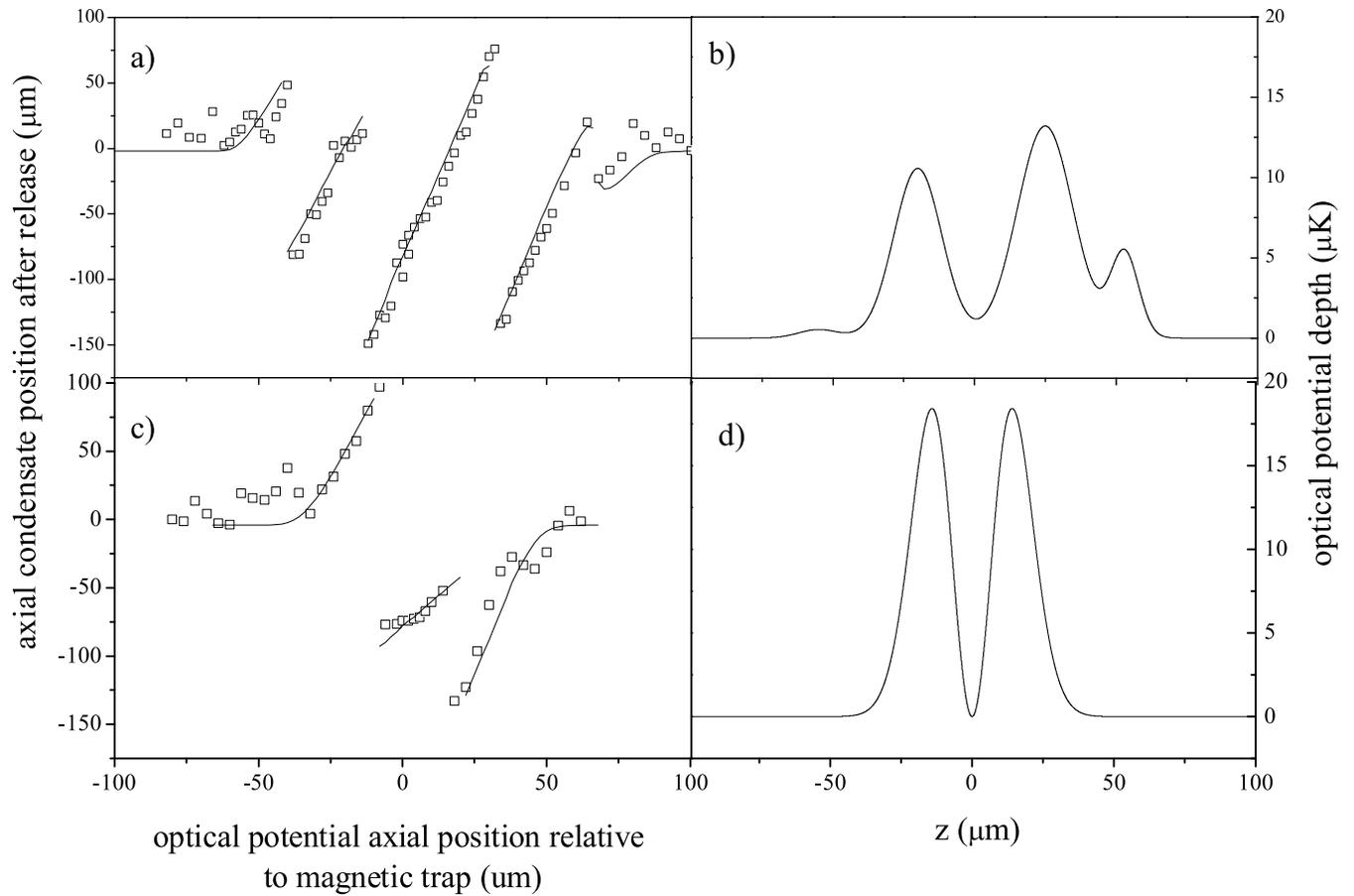}\\
\caption{The position of the dipole trap beam is scanned in the
$z$ direction, while the magnetic trap defining the initial
position of the atoms, remains fixed. The dipole trap intensity is
then ramped on and the position of the atoms observed after 15~ms
of free expansion. (a)/(c) show a scan far from/closer to the
focus of the trapping beam, `white squares' are experimental
points and the `solid line' is the deflection expected for the
potential in (b)/(d).\label{alignmentandprofile}}
\end{figure}

In Fig.~\ref{alignmentandprofile} the position of the dipole trap
beam is scanned in the $z$ direction, while the magnetic trap,
which defines the initial position of the atoms, remains fixed.
The dipole trap intensity is then ramped on and the position of
the atoms observed after free expansion.
Fig.~\ref{alignmentandprofile}(c) shows the pattern observed and
expected for a TEM$_{01}$ potential, as shown in
Fig.~\ref{alignmentandprofile}(d). The magnetic and dipole traps
are only aligned in the central region of the plot in
Fig.~\ref{alignmentandprofile}(c), this coincides with a sharp
increase in the axial expansion of the condensate. The beam waist
can be accurately measured from the width of the central region.
This method can also be used to profile the beam at various points
along its optical axis, and to partially reconstruct the intensity
profile. Fig.~\ref{alignmentandprofile}(a) shows a scan further
away from the beam focus, and Fig.~\ref{alignmentandprofile}(b) is
an intensity pattern that produces the observed deflection.

\subsection{Loading the Trap}

Our experiment uses evaporative cooling in a TOP trap to create a
Bose-condensate that contains $\sim10^{5}$ atoms of the Rb$^{87}$
isotope in the $|F=1, m_{F}=-1 \rangle$ state. The initial
oscillation frequencies of atoms in the magnetic trap are 61~Hz
radially and 172.5~Hz along the $z$ (axial) direction. After a
condensate has been formed it is adiabatically loaded into the
combined potential: the intensity of the dipole force laser beam
is ramped from zero to its maximum value in 100-300 ms (depending
on the strength of the optical potential). This ensures that no
dipole modes, or internal modes of the condensate, are excited
during the loading phase. This was confirmed by an exact numerical
simulation. The combined central position changes most rapidly
when the frequencies of the potentials are comparable (see
Eq.~\ref{zposeqn}), at the beginning of the ramp. For the highest
frequency dipole traps this means that the ramp needs to be
extended to give the condensate a smooth transition and avoid
unnecessary heating.

\subsection{Calibration}

\begin{figure}
\includegraphics[width=\columnwidth]{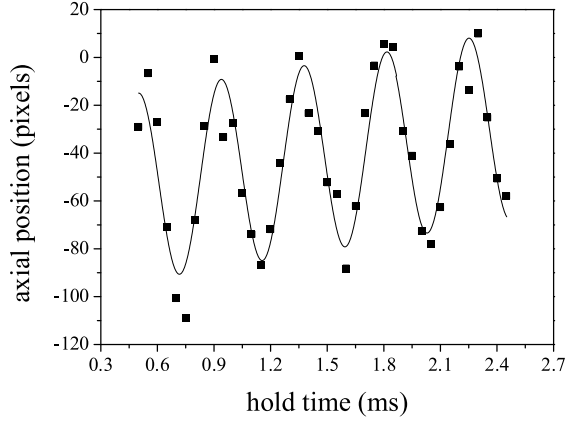}\\
\caption{Axial frequency measurement with TOP field turned off.
`Black squares' are experimental points, `solid line' is a
sinusoidal fit to the data used to extract the
frequency.\label{maxfrequencymeasurement}}
\end{figure}

\begin{figure}
\includegraphics[width=\columnwidth]{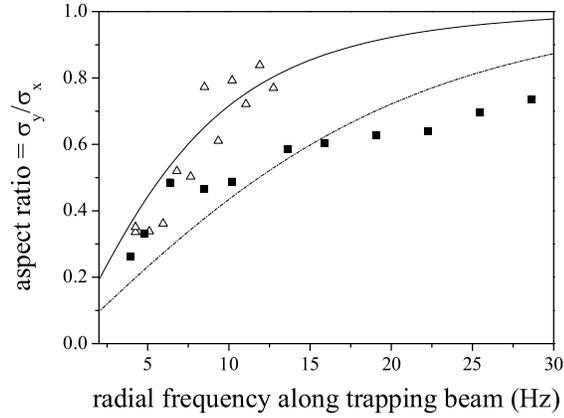}\\
\caption{Aspect ratio of the condensate, defined as
$\sigma_{y}/\sigma_{x}$, after 20~ms free expansion. `White
triangles' and `black squares' are the measured aspect ratios of a
cloud released from a hybrid trap with $f_{z}$=960~Hz and $f_{z}$
= 1990~Hz respectively. `Solid line' and `dashed line' are fits to
the data using the variational method; this allows the frequency
of the optical potential along $y$ to be determined. For the data
given $f_{y}$=12.0~Hz and $f_{y}$=26.0~Hz just from the optical
potential, as expected there is more confinement in the trap with
the higher axial frequency.\label{aspectratioplot}}
\end{figure}

Dipole frequency measurements in the radial and axial directions
are used to characterise the trap. For the highest frequency
traps, the measurement is complicated because the TOP frequency
(7~kHz) is comparable to the dipole frequency in the axial
direction. Efforts to observe the dipole oscillation with the TOP
field on were unsuccessful for this reason. It is therefore
necessary to turn off the TOP field and move the quadrupole center
above the optical potential, to avoid atom loss through Majorana
spin flips. In this geometry it is necessary to suddenly change
the quadrupole gradient to excite the dipole mode. We have
measured an axial frequency of 2.2~kHz, as shown in
Fig.~\ref{maxfrequencymeasurement}, for a beam waist of
7.2~$\mu$m, determined from the deflection of atoms as previously
explained. This agrees roughly with the expected value of 2.7~kHz
for an ideal phase-plate profile, calculated using
Eq.~\ref{Omega1}, when the rotational misalignment between the
phase-plate and the final cylindrical lens is taken into account.
A rotation by only $4\pm1$ degrees effectively changes the axial
frequency to $1.94 \mp 0.27~\text{kHz}$. From the frequency
measurement it is also possible to estimate the axial beam drift
to be approximately 18~$\mu$m/hr. The radial trap frequencies have
also been measured by exciting dipole modes, and by measuring the
aspect ratio of the condensate (defined as $l_y/l_x$, where $l$ is
the width of the condensate extracted from a Gaussian fit) in the
$xy$ plane after free expansion. At high values of radial
confinement from the magnetic trap, the dipole frequencies were as
expected for a purely magnetic potential; however, as the magnetic
confinement was reduced it was observed that the cloud elongated
along the beam direction (see Fig.~\ref{aspectratioplot}), and
when the dipole mode was excited it oscillated only along the
propagation direction of the trapping beam. In this direction the
dipole oscillation frequency agrees with the expected value for a
purely magnetic potential, but across the beam there is some extra
confinement provided by the optical potential. The optical
frequency along the $y$ direction was used as a fitting parameter
for the data in Fig.~\ref{aspectratioplot}: for a fixed optical
frequency a curve was calculated for all the various magnetic
frequencies and the resulting curve compared to the data points.
This procedure was repeated to find the least squares minimum
value for the optical potential. This was found to be equivalent
to 26~Hz for an axial trapping frequency of 1990~Hz, and is caused
by the rotational misalignment described above. Indeed
subsequently we have measured the tilt of the input beam relative
to the final cylindrical lens and found a value of $4\pm1$
degrees, from Fig.~\ref{anglemisaligncombined} it can be seen that
this theoretically gives between 19 and $23$~Hz of confinement.

\begin{figure}
\includegraphics[width=\columnwidth]{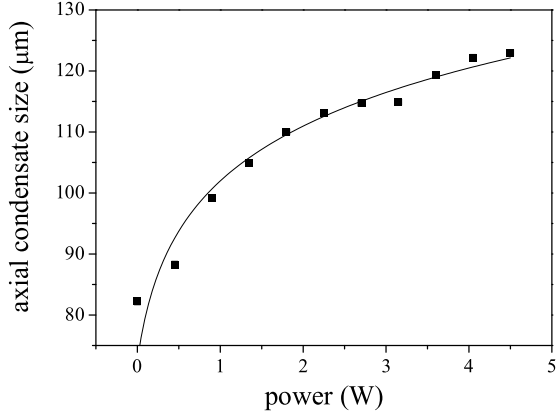}\\
\caption{Axial condensate size after 15~ms of free expansion, as a
function of power. The axial/radial trap frequency is
960~Hz/61~Hz. `Black squares' are the experimental data, `black
line' is the hydrodynamic prediction provided that
$f_{z}\propto\sqrt{P}$, as is expected for a harmonic
potential.\label{expansionvspower}}
\end{figure}

We also measured the axial size of the condensate as a function of
trapping beam power. The condensate expanded for 15~ms before it
was imaged. The results are shown in Fig.~\ref{expansionvspower}
and are in good agreement with the hydrodynamic theory and the
assumption that $\Omega\propto\sqrt{P}$, as would be expected for
a harmonic dipole trap (see Eq.~\ref{Omega1}). These measurements
were made at a radial frequency of 61~Hz and so the hydrodynamic
expansion theory applies.

\section{The Quasi-2D Regime}

\subsection{Criterion for Q2D}

The condition $\mu<\hbar\omega_z$, where $\mu$ is the chemical
potential, can be used as a criterion for when a condensate is in
the Q2D regime. In this case the ground state energy is smaller
than the harmonic oscillator spacing and thus comparable to the
harmonic oscillator ground state. When $\sim10^{5}$ atoms have
been loaded into our combined optical and magnetic potential the
Q2D regime can be reached by either decreasing the number of
atoms, or by adiabatically decreasing the radial confinement
provided by the magnetic trap. In the following experiments we use
the second approach, to provide more control. Along the $z$
direction the condensate shape becomes very similar to the
Gaussian profile of an ideal gas. However, along the weakly
confined $x$ and $y$ axes the condensate is characterised by the
hydrodynamic parabolic shape. For the best description in terms of
simple analytic functions it is therefore best to describe the
condensate wavefunction as a hybrid of a parabolic and a Gaussian
distrubution.

\subsection{Expansion of a Q2D gas}

\begin{figure}
\includegraphics[width=\columnwidth]{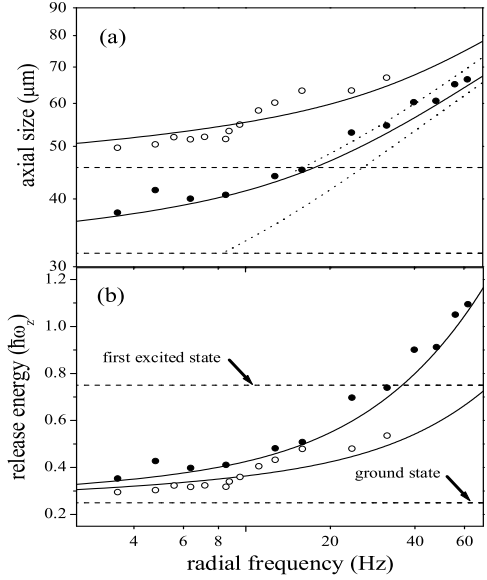}\\
\caption{(a) The axial expansion of the condensate after 15~ms of
free expansion for various trap geometries and atom numbers. Solid
lines indicate theoretical variational predictions, dashed lines
indicate the ideal gas limit and dotted lines the hydrodynamic
limit. The data are taken for traps with $\omega_z/2\pi=1990$~Hz
(open circles) and 960~Hz (filled circles). The atom numbers are
$8\times 10^4$ and $1.1\times 10^5$, respectively. (b) The release
energy of the condensate derived from the expansion measurements
in (a). The energy tends towards the vertical zero-point kinetic
energy as $\omega_{x}$ is reduced.
\label{expansionvsradialfrequency}}
\end{figure}

In our experiments we observed the condensate expansion in various
regimes and the smooth transition from the hydrodynamic expansion
characteristics \cite{Castin1996a} to those of a quasi-2D gas,
which essentially expands like an ideal non-interacting gas in the
direction of tight confinement. The results are shown in
Fig.~\ref{expansionvsradialfrequency}(a), where the open circles
and filled circles are the data for traps with
$\omega_z/2\pi=1990~ \text{Hz}$ and 960~Hz respectively. The
expansion time is constant at 15~ms and the radial trap frequency
is varied to explore the transition to Q2D. These curves
demonstrate clearly the ideal-gas like behaviour for the axial
expansion in this limit. We compared our findings to those of
theoretical predictions which were derived from variational models
\cite{Perez-Garcia1996a} and found good agreement to our data. The
prediction of the hybrid variational model
\cite{Hechenblaikner2004a}, which is indistinguishable from that
of the Gaussian variational model, is given by the solid line. The
horizontal dashed lines indicate the expansion of the ideal gas,
given by $R_{z}=t\sqrt{\hbar\omega_{z}/m}$, where $t$ is the
expansion time, and the dotted lines the expansion of the
hydrodynamic gas. The `ideal gas' and hydrodynamic models yield
straight lines on the logarithmic plot. In contrast, the curve for
the hybrid model follows the hydrodynamic asymptote down to
$\omega_x\approx20~\text{Hz}$, corresponding to
$\mu\approx\hbar\omega_z$, where it bends and follows the `ideal
gas' line towards zero radial frequency. The two noticeably do not
coincide at $\omega_x\approx 0$ because of the residual optical
anisotropy in the radial plane (as we discussed before in the text
describing Fig.~\ref{aspectratioplot}), which was taken into
account in our variational model and is in good agreement with the
experimental data. This transition from the hydrodynamic to the
`ideal gas' asymptote gives conclusive evidence of the gas
entering the Q2D regime. The release energy of the condensate,
calculated from the size after expansion, is also shown in
Fig.~\ref{expansionvsradialfrequency}(b): this tends towards the
ideal gas limit of $\hbar\omega_{z}/4$ as $\omega_{x}$ is reduced.
This is the vertical kinetic zero-point energy; all that is
available if the confining potential is suddenly switched off.

\section{Conclusions and outlook}
In these experiments we have studied the properties of quasi-2D
condensates. To obtain the required geometry we increased the
stiffness of the axial confinement by superimposing an optical
potential on the existing magnetic trapping potential. The
resulting trap was characterised in detail and we obtained
confinement frequencies close to the theoretically expected
values. We then studied the properties of Q2D condensates
including their chemical potential and expansion characteristics.
Our experimental observations agreed with theoretical predictions
to confirm that we successfully entered the Q2D regime. This was
the first experiment to explore the hydrodynamic-Q2D transition by
gradually increasing the trap anisotropy instead of throwing atoms
away as in previous work \cite{Gorlitz2001}. In future experiments
we hope to study the properties of vortices and vortex arrays in
the Q2D regime and investigate the possibilities of observing a
Berezinskii-Kosterlitz-Thouless transition and studying its
properties.

\begin{acknowledgments}
The authors would like to thank Mr. Chris Goodwin for designing
and manufacturing the phase-plate using the Thin Film Facility at
the Oxford Physics Department. The authors would also like to
acknowledge financial support from the EPSRC and DARPA.
\end{acknowledgments}

\bibliography{intro}

\clearpage

\end{document}